\documentclass[twoside,11pt]{article}
\usepackage{pgm}
\usepackage[utf8]{inputenc}
\inputencoding{utf8}
\usepackage{lineno}
\usepackage{hyperref,color,soul,booktabs}
\setulcolor{blue}
\newcommand{\BibTeX}{\textsc{B\kern-0.1emi\kern-0.017emb}\kern-0.15em\TeX}
\usepackage{pifont}
\usepackage{amssymb}
\usepackage{colortbl}

\usepackage{multirow}
\usepackage{booktabs,amsfonts,dcolumn}
\usepackage{graphicx}
\usepackage{amsmath}
\usepackage{geometry}
\usepackage{algorithmic}
\usepackage{pgf, tikz}
\usetikzlibrary{arrows,automata,fit}
\usetikzlibrary{shapes}

\ShortHeadings{Shooting performance in professional biathlon}{M. Leonelli}

\begin{document}
\title{Predicting and understanding shooting performance in professional biathlon: A Bayesian approach}
\author{
   \Name{Manuele Leonelli} \Email{manuele.leonelli@ie.edu}\\
   \addr School of Science and Technology, IE University, Madrid, Spain}


\maketitle

\begin{abstract} Biathlon is a unique winter sport that combines precision rifle marksmanship with the endurance demands of cross-country skiing. We develop a Bayesian hierarchical model to predict and understand shooting performance using data from the 2021/22 Women’s World Cup season. The model captures athlete-specific, position-specific, race-type, and stage-dependent effects, providing a comprehensive view of shooting accuracy variability. By incorporating dynamic components, we reveal how performance evolves over the season, with model validation showing strong predictive ability at both overall and individual levels. Our findings highlight substantial athlete-specific differences and underscore the value of personalized performance analysis for optimizing coaching strategies. This work demonstrates the potential of advanced Bayesian modeling in sports analytics, paving the way for future research in biathlon and similar sports requiring the integration of technical and endurance skills. \end{abstract}


\section{Introduction}
\label{sec:introduction}

Biathlon has been an Olympic winter sport since the 1960 games of Squaw Valley, uniquely combining the precision of rifle marksmanship with the endurance and strength of cross-country skiing while carrying a rifle. This dual nature of biathlon demands a unique blend of physiological, psychological, and technical skills, making it one of the most demanding winter sports. With the sport's growing popularity, the number of countries represented in the World Cup has increased steadily, indicating a widening global interest.

Biathlon competitions include several race formats, each affecting an athlete's shooting and skiing strategies. The \textit{sprint} race, covering 7.5 km for women, has two shooting bouts (one prone and one standing), with each miss incurring a 150-meter penalty loop. The \textit{pursuit} race (10 km) includes four shooting bouts, with athletes starting staggered based on previous sprint results. The \textit{individual} race, the longest at 15 km, features four shooting bouts and a one-minute time penalty per miss, emphasizing shooting accuracy. Finally, the \textit{mass start} (12.5 km) has all athletes starting together, with four shooting bouts and the same 150-meter penalty loop per miss. In all formats, athletes shoot half in the prone position and half standing. Prone shooting generally yields higher hit rates due to greater stability, while standing is more challenging.

Previous research by \citet{maier2018predicting} has shown that shooting scores significantly impact race rankings, with several factors affecting shooting accuracy across race types and positions. Lower hit rates are observed in the sprint and pursuit disciplines compared to individual and mass start races, and standing shooting yields lower accuracy than prone shooting. Furthermore, within each shooting bout, certain shots—such as the first prone and fifth standing shots—tend to have the lowest accuracy. Physiological and biomechanical factors, including low vertical rifle motion in prone shooting and reduced body sway in standing, are essential for shooting success \citep{sattlecker2014,sattlecker2017}. Elevated heart rates and accumulated fatigue from skiing also impact accuracy \citep{groslambert1999validation,hoffman1992characterization}.

Research also indicates that an athlete’s preceding mode-specific hit rate is a dominant predictor of shooting performance, though considerable randomness persists \citep{maier2018predicting}. Top-ranked athletes have demonstrated higher hit rates, and significant within-athlete variability has been observed in accuracy across shots \citep{luchsinger2018comparison}. Contextual factors, such as crowd presence and competitor proximity, also influence performance. For instance, supportive crowds impact shooting and skiing outcomes differently based on an athlete's rank and gender \citep{harb2019choking,heinrich2021selection}.

While data analytics studies on shooting performance in biathlon are extensive (e.g., \citet{heinrich2022impact,maier2018predicting,skattebo2018variability}), existing methods provide valuable insights but have limited capacity for capturing the complex, nested structure of biathlon performance data. To address these challenges, \textit{Bayesian hierarchical models} \citep{gelman2013bayesian} offer a flexible framework for capturing complex, nested, and probabilistic relationships, accommodating both individual and group-level variability to provide a deeper understanding of performance factors. Bayesian methods are gaining traction in sports analytics, where they offer flexibility in handling complex data structures and provide probabilistic insights into performance factors \citep{santos2019bayesian}. Bayesian models have been applied in various sports, including basketball \citep{hobbs2020bayesian}, football \citep{baio2010bayesian,tsokos2019modeling}, swimming \citep{wu2022bayesian}, tennis \citep{ingram2019point}, and volleyball \citep{egidi2020,gabrio2021bayesian}. However, biathlon shooting performance remains largely unexplored with Bayesian techniques, with only one example in winter sports analytics overall \citep{glickman2015stochastic}. Furthermore, \textit{dynamic components} are incorporated to account for stage-specific effects across the season, acknowledging that shooting accuracy may fluctuate over time due to training, adaptation, or fatigue.

This study aims to apply a Bayesian approach to better understand and predict shooting performance in professional biathlon. By analyzing data from the 2021/22 Women’s World Cup season - chosen for its increased number of races, which provides a more robust dataset for analysis - this model seeks to reveal the nuanced factors affecting shooting accuracy and quantify the predictive power of these variables. Importantly, the model uses a simplified structure that focuses on the most basic components of a World Cup season. This approach not only allows for quick and efficient estimation, but also ensures that the model can be easily applied to different seasons, genders, or youth races using publicly available data, making it broadly useful for athletes and coaches.

\section{Materials and Methods}

\subsection{Data and Design}
The 2021/22 Women’s World Cup season consisted of 26 races: 10 sprints, 3 individual races, 8 pursuits, and 5 mass start events. Data for this study was sourced from the IBU Datacenter\footnote{\href{https://www.biathlonresults.com}{https://www.biathlonresults.com}}, covering this season, including the 2022 Beijing Winter Olympics. This specific season was selected due to its increased number of races, which provides a larger sample size and allows for more accurate conclusions. Our analysis focuses on the top 30 female athletes based on the end-of-season rankings, ensuring that the dataset includes athletes with consistent participation and high-level performance.

For each shooting session, the primary outcome variable was the number of hits recorded (an integer between zero and five). Additional predictor variables included:

\begin{itemize} \item \textbf{Shooting Position}: Prone or standing, as each race requires athletes to alternate between the two. \item \textbf{Race Type}: One of four formats—individual, sprint, pursuit, or mass start—each with different scoring structures and participation requirements. \item \textbf{World Cup Stage}: An integer between one and eleven, as the season included eleven stages. \end{itemize}

The final dataset consists of 2,088 observations, as not all athletes participated in every event. For instance, Russian and Belarusian athletes were absent from the final three stages due to external restrictions. Additionally, certain race formats impose restrictions on participation based on previous race results or rankings: athletes who finished outside the top 60 in a sprint event did not qualify for the following pursuit, and only 30 athletes participated in mass start events. Mass start selection is based partially on current standings, with the top-ranked athletes qualifying directly, while additional athletes qualify based on their stage performance. For the Olympic events, additional eligibility criteria were in place, such as a cap of four athletes per country.

Despite the availability of more granular data, our model focuses on easily accessible and fundamental components like shooting position, race type, and stage effects. This simplification provides a practical model that can be quickly estimated and applied, offering an efficient way to profile athletes and understand performance dynamics. Moreover, this approach enables broad applicability, as similar data can be effortlessly scraped from public resources, facilitating its use for other seasons, different genders, or youth competitions.

\subsection{Exploratory Data Analysis}

Before modeling, we conducted an exploratory analysis to identify patterns in shooting performance and potential predictors. We summarized shooting accuracy by position (prone and standing) and race type (individual, sprint, pursuit, and mass start), which highlighted notable performance differences across these conditions. To visualize consistency and fluctuations throughout the season, we created heatmaps showing each athlete’s accuracy across the eleven World Cup stages, organized by stage order and color-coded to illustrate performance trends over time.

We further analyzed shooting patterns by applying hierarchical clustering, which grouped athletes based on similarities in their shooting profiles across positions and race types, revealing distinct group-specific trends. Additionally, we evaluated the association between end-of-season rankings and shooting accuracy using Spearman’s rank correlation. These insights informed the structuring of our Bayesian hierarchical model and guided the selection of priors to effectively capture the complexity of biathlon performance.

\subsection{The Model}

In this study, informed by the exploratory analysis, we employ a Bayesian hierarchical model with dynamic stage effects, which allows for time-varying performance across the season while maintaining stable athlete-specific skills across shooting positions and race types. This model is intuitively aligned with the structure of biathlon competitions, where we expect stage-to-stage variations in performance, while each athlete’s shooting skill and race-specific behavior remain relatively stable. This model thus provides a straightforward yet effective framework for analyzing biathlon shooting performance.

Let $Y_i$ denote the shooting outcome for the $i$-th session, recorded as the number of hits (an integer between 0 and 5). The outcome is assumed to follow a Binomial distribution:
\[
Y_i \sim \text{Binomial}(5, p_i)
\]
where $p_i$ represents the probability of hitting a target during the $i$-th session. We model $p_i$ on the logit scale to incorporate various athlete-specific effects:
\[
\text{logit}(p_i) = \mu_{t[i]} + \beta_{s[i], t[i]} + \gamma_{s[i], x[i]} + \omega_{s[i], z[i]}
\]
where:
\begin{itemize}
\item $\mu_{t[i]}$ denotes the baseline log-odds of hitting a target at stage $t$,
\item $\beta_{s[i], t[i]}$ is a dynamic, athlete-specific effect that varies across stages $t$,
\item $\gamma_{s[i], x[i]}$ is an athlete-specific skill effect for each shooting position $x$ (prone or standing),
\item $\omega_{s[i], z[i]}$ represents the athlete-specific interaction with each race type $z$ (individual, sprint, pursuit, or mass start).
\end{itemize}
Subscripts in square brackets indicate indexing for variables specific to each observation $i$.

In a Bayesian modeling approach, the specification of prior distributions completes the model, combining with the likelihood to yield the posterior distributions of the parameters. This allows us to incorporate prior knowledge or assumptions about parameter values into the analysis.

We adopt a Markov structure for the stage-specific effects, capturing the temporal dynamics in shooting performance across the season. The priors for these effects are specified as follows:
\[
\mu_1 \sim \mathcal{N}(0, \tau_\mu), \quad \mu_t \sim \mathcal{N}(\mu_{t-1}, \tau_\mu) \text{ for } t > 1,
\]
where the $\mu_t$ parameters evolve over time, forming a dynamic linear model that captures changes in baseline performance from stage to stage. This structure acknowledges that stage-specific performance is likely correlated across consecutive stages.

For the athlete-specific dynamic effects, we similarly define:
\[
\beta_{s, 1} \sim \mathcal{N}(0, \tau_\beta), \quad \beta_{s, t} \sim \mathcal{N}(\beta_{s, t-1}, \tau_\beta) \text{ for } t > 1.
\]
This Markov structure models how each athlete’s performance evolves over the season, allowing for smooth, stage-to-stage variations while reflecting each athlete’s individual trajectory.

The position-specific and race-type effects for each athlete are modeled as:
\[
\gamma_{s, x} \sim \mathcal{N}(0, \tau_\gamma), \quad \omega_{s, z} \sim \mathcal{N}(0, \tau_\omega).
\]
These priors are specified independently, without a temporal component, reflecting the assumption that an athlete's skill differences across shooting positions and race types are stable throughout the season.

To ensure identifiability, we impose constraints on the model parameters. For the athlete-stage effects $\beta_{s[i], t[i]}$, we use a sum-to-zero constraint:
\[
\sum_{s=1}^{N_{\text{athletes}}} \beta_{s, t} = 0 \quad \text{for each stage } t.
\]
For shooting position and race type, we apply similar constraints separately for each athlete to highlight individual skill differences:
\[
\sum_{x=1}^{2} \gamma_{s, x} = 0 \quad \text{and} \quad \sum_{z=1}^{4} \omega_{s, z} = 0.
\]
These choices ensure the interpretability of parameters by clearly capturing each athlete's relative strengths and weaknesses across positions and race formats.

By incorporating this structure, the model effectively balances the need for flexibility with the interpretability of temporal and athlete-specific effects, capturing the dynamic nature of biathlon performance across the season. While we considered simpler models with fewer parameters, such as those without stage-specific dynamics, and explored more complex models with dynamic effects for position and race, the presented model provides the right balance between quality of fit and interpretability. By capturing only stage-to-stage variations in performance, the model aligns with expected biathlon dynamics, reflecting how athletes’ shooting performance fluctuates over time while maintaining stable position- and race-specific skills. This choice offers an effective framework for analyzing biathlon shooting performance with an optimal trade-off between model fit and simplicity.

\subsection{Model Estimation and Implementation}

The models were implemented in \texttt{R} using \texttt{RStudio} and estimated with \texttt{JAGS} through the \texttt{rjags} interface \citep{plummer2003jags}. Each model was run with four Markov chains, using a burn-in period of 1000 iterations to ensure convergence. After the burn-in, we retained 5000 iterations per chain, applying a thinning factor of 5 to reduce autocorrelation. This resulted in a total posterior sample size of 4000 when combining the four chains.

To verify model convergence and ensure reliable parameter estimates, we monitored standard diagnostics such as trace plots and effective sample sizes. These diagnostics confirmed that our chosen sampling settings provided a suitable balance between computational efficiency and the need for robust posterior estimation.

For full reproducibility, all code for data preparation, model specification, and analysis is available on GitHub.

\subsection{Bayesian Predictions}

A key advantage of the Bayesian framework is its ability to generate probabilistic predictions by drawing on the posterior distribution of the model parameters. Once the model is fitted, we obtain a posterior sample of the parameters, which captures the uncertainty associated with each effect in the model. To make predictions, we use these posterior samples to calculate the predicted probability of hitting a target for each athlete, shooting position, race type, and stage. This process is straightforward: we simulate outcomes by sampling from the predictive distribution, reflecting both parameter uncertainty and the inherent variability in the shooting data.

This approach also provides prediction intervals, offering a clear measure of uncertainty around each prediction. For instance, by summarizing the posterior predictive distribution (e.g., calculating the 95\% credible interval), we can quantify the expected range of shooting outcomes. This is particularly useful for coaches and performance analysts, as it not only forecasts potential performance but also highlights the variability and risks associated with each prediction, enabling more informed decision-making in training and competition strategy.

\section{Results}

\subsection{Exploratory Analysis of Shooting Performance}

\begin{table}
\centering
\caption{Biathlon performance summary by shooting position and race format, including overall accuracy for the season. \label{table}}
\centering
\scalebox{0.65}{
\begin{tabular}[t]{lccccccccc}
\toprule
\multicolumn{1}{c}{ } & \multicolumn{2}{c}{\textbf{Position}} & \multicolumn{4}{c}{\textbf{Race}} & \multicolumn{3}{c}{\textbf{Overall Totals}} \\
\cmidrule(l{3pt}r{3pt}){2-3} \cmidrule(l{3pt}r{3pt}){4-7} \cmidrule(l{3pt}r{3pt}){8-10}
\textbf{Athlete} & \textbf{Prone} & \textbf{Standing} & \textbf{Individual} & \textbf{Sprint} & \textbf{Pursuit} & \textbf{Mass Start} & \textbf{Total Shots} & \textbf{Total Hits} & \textbf{Overall}\\
\midrule
M. Olsbu Roiseland & 92.6\% & 88.9\% & 87.5\% & 92.0\% & 91.9\% & 88.8\% & 380 & 345 & 90.8\%\\
E. Oeberg & 82.6\% & 87.9\% & 80.0\% & 82.0\% & 86.9\% & 88.8\% & 380 & 324 & 85.3\%\\
L.T. Hauser & 90.0\% & 86.2\% & 88.3\% & 90.0\% & 88.8\% & 85.0\% & 420 & 370 & 88.1\%\\
H. Oeberg & 79.5\% & 80.5\% & 75.0\% & 80.0\% & 81.2\% & 80.0\% & 380 & 304 & 80.0\%\\
A. Chevalier Bouchet & 88.4\% & 82.1\% & 85.0\% & 85.0\% & 88.6\% & 81.0\% & 380 & 324 & 85.3\%\\
\addlinespace
D. Herrmann Wick & 85.6\% & 81.7\% & 95.0\% & 83.0\% & 80.7\% & 83.8\% & 360 & 301 & 83.6\%\\
D. Alimbekava & 92.9\% & 82.6\% & 90.0\% & 92.9\% & 87.5\% & 80.0\% & 310 & 272 & 87.7\%\\
J. Braisaz Bouchet & 73.5\% & 79.0\% & 81.7\% & 71.0\% & 73.6\% & 82.0\% & 400 & 305 & 76.2\%\\
D. Wierer & 87.0\% & 82.5\% & 83.3\% & 82.0\% & 87.1\% & 85.0\% & 400 & 339 & 84.8\%\\
M. Davidova & 85.5\% & 84.0\% & 93.3\% & 79.0\% & 85.0\% & 85.0\% & 400 & 339 & 84.8\%\\
\addlinespace
T. Eckhoff & 88.5\% & 76.4\% & 76.7\% & 83.3\% & 83.0\% & 85.0\% & 330 & 272 & 82.4\%\\
J. Simon & 85.5\% & 77.5\% & 80.0\% & 73.0\% & 87.1\% & 83.0\% & 400 & 326 & 81.5\%\\
V. Voigt & 91.0\% & 92.0\% & 95.0\% & 95.0\% & 88.8\% & 91.0\% & 400 & 366 & 91.5\%\\
A. Bescond & 85.5\% & 76.5\% & 75.0\% & 86.0\% & 86.2\% & 68.8\% & 400 & 324 & 81.0\%\\
I.L. Tandrevold & 91.9\% & 83.2\% & 88.3\% & 83.3\% & 90.0\% & 87.5\% & 370 & 324 & 87.6\%\\
\addlinespace
M. Brorsson & 92.9\% & 81.2\% & 91.7\% & 86.2\% & 85.8\% & 86.2\% & 340 & 296 & 87.1\%\\
J. Jislova & 92.5\% & 89.5\% & 86.7\% & 94.0\% & 90.6\% & 91.2\% & 400 & 364 & 91.0\%\\
H. Sola & 80.0\% & 75.2\% & 71.7\% & 81.4\% & 83.0\% & 70.0\% & 290 & 225 & 77.6\%\\
L. Persson & 86.5\% & 80.0\% & 80.0\% & 84.0\% & 84.3\% & 81.7\% & 340 & 283 & 83.2\%\\
K. Reztsova & 80.0\% & 80.0\% & 92.5\% & 81.4\% & 73.3\% & 83.3\% & 290 & 232 & 80.0\%\\
\addlinespace
F. Preuss & 89.6\% & 82.2\% & 80.0\% & 78.6\% & 92.0\% & 88.3\% & 270 & 232 & 85.9\%\\
C. Chevalier & 82.6\% & 81.3\% & 80.0\% & 81.1\% & 79.2\% & 90.0\% & 310 & 254 & 81.9\%\\
P. Batovska Fialkova & 81.8\% & 76.5\% & 83.3\% & 73.0\% & 85.0\% & 76.2\% & 340 & 269 & 79.1\%\\
F. Hildebrand & 95.4\% & 84.6\% & 87.5\% & 88.3\% & 93.0\% & 88.3\% & 260 & 234 & 90.0\%\\
U. Nigmatullina & 84.1\% & 86.2\% & 78.3\% & 82.9\% & 90.0\% & 86.7\% & 290 & 247 & 85.2\%\\
\addlinespace
V. Hinz & 92.1\% & 80.0\% & 92.5\% & 83.3\% & 87.1\% & 83.3\% & 330 & 284 & 86.1\%\\
M. Eder & 78.9\% & 71.6\% & 85.0\% & 75.0\% & 71.4\% & 75.0\% & 380 & 286 & 75.3\%\\
L. Lie & 91.4\% & 89.1\% & 86.7\% & 84.4\% & 92.9\% & 96.7\% & 350 & 316 & 90.3\%\\
K.O. Knotten & 88.6\% & 83.6\% & 90.0\% & 93.8\% & 81.7\% & 80.0\% & 280 & 241 & 86.1\%\\
S. Nilsson & 78.5\% & 67.7\% & 80.0\% & 73.8\% & 71.0\% & 73.3\% & 260 & 190 & 73.1\%\\
\addlinespace
\textbf{Overall} & \textbf{86.5\%} & \textbf{81.9\%} & \textbf{84.5\%} & \textbf{83.2\%} & \textbf{85.1\%} & \textbf{83.6\%} & \textbf{10440} & \textbf{8788} & \textbf{84.2\%}\\
\bottomrule
\end{tabular}}
\end{table}

The overall shooting accuracy for the season stood at 84.2\%, reflecting a generally strong performance among top athletes. Table~\ref{table} provides a summary of shooting accuracy across different shooting positions and race formats. There was a notable difference between shooting positions, with athletes achieving an average accuracy of 86.5\% in the prone position compared to 81.9\% in the standing position, highlighting the increased stability and accuracy associated with prone shooting. Across different race formats, the accuracy rates remained relatively consistent, with the pursuit race having the highest average accuracy at 85.1\% and the sprint race the lowest at 83.2\%.

This analysis also indicated athlete-specific shooting characteristics. For instance, a majority of the athletes (24 out of the top contenders) performed better in the prone position than in standing. Examining race-specific performance, the pursuit format emerged as the highest-performing race for 12 athletes, while the individual format followed closely. Conversely, the sprint race was noted as the most challenging format, being the lowest-performing race for 9 athletes, as shown in Table~\ref{table1}.

\begin{table}
\centering
\caption{Counts of athletes for each race format as favorite and least favorite in terms of shooting accuracy. \label{table1}}
\scalebox{0.65}{
\begin{tabular}{lcc}
\toprule
\textbf{Race Format} & \textbf{Favorite Race Count} & \textbf{Least Favorite Race Count} \\
\midrule
Individual           & 8                             & 8                                  \\
Mass Start           & 5                             & 6                                  \\
Pursuit              & 12                            & 7                                  \\
Sprint               & 5                             & 9                                  \\
\bottomrule
\end{tabular}}
\end{table}

\begin{figure}
    \centering
    \includegraphics[width=0.8\textwidth]{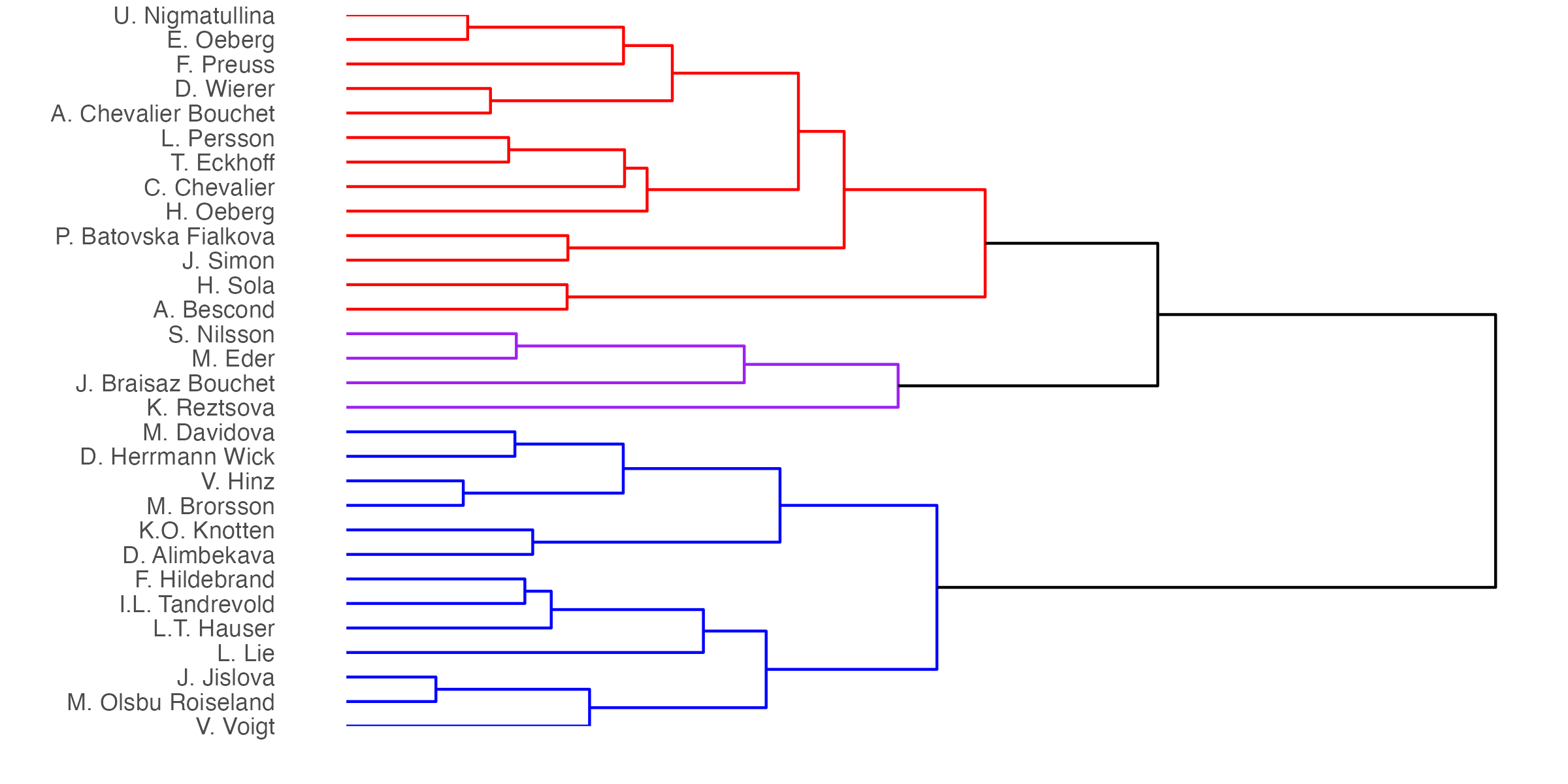}
    \caption{Hierarchical clustering dendrogram showing three distinct athlete clusters based on shooting accuracy profiles across shooting positions and race types.}
    \label{fig:dendrogram}
\end{figure}

The hierarchical clustering analysis (Figure~\ref{fig:dendrogram}) uncovered three main groups of athletes based on shooting accuracy across positions and race types. The first cluster consisted of high-performing athletes with above-average accuracy across all races, particularly in individual and sprint formats. The second cluster, with moderate performance, showed slightly lower accuracy in individual races compared to other race types. The third cluster included athletes with lower overall accuracy, though individual races were where they performed best despite their generally lower scores.

\begin{figure}
    \centering
    \includegraphics[width=0.9\textwidth]{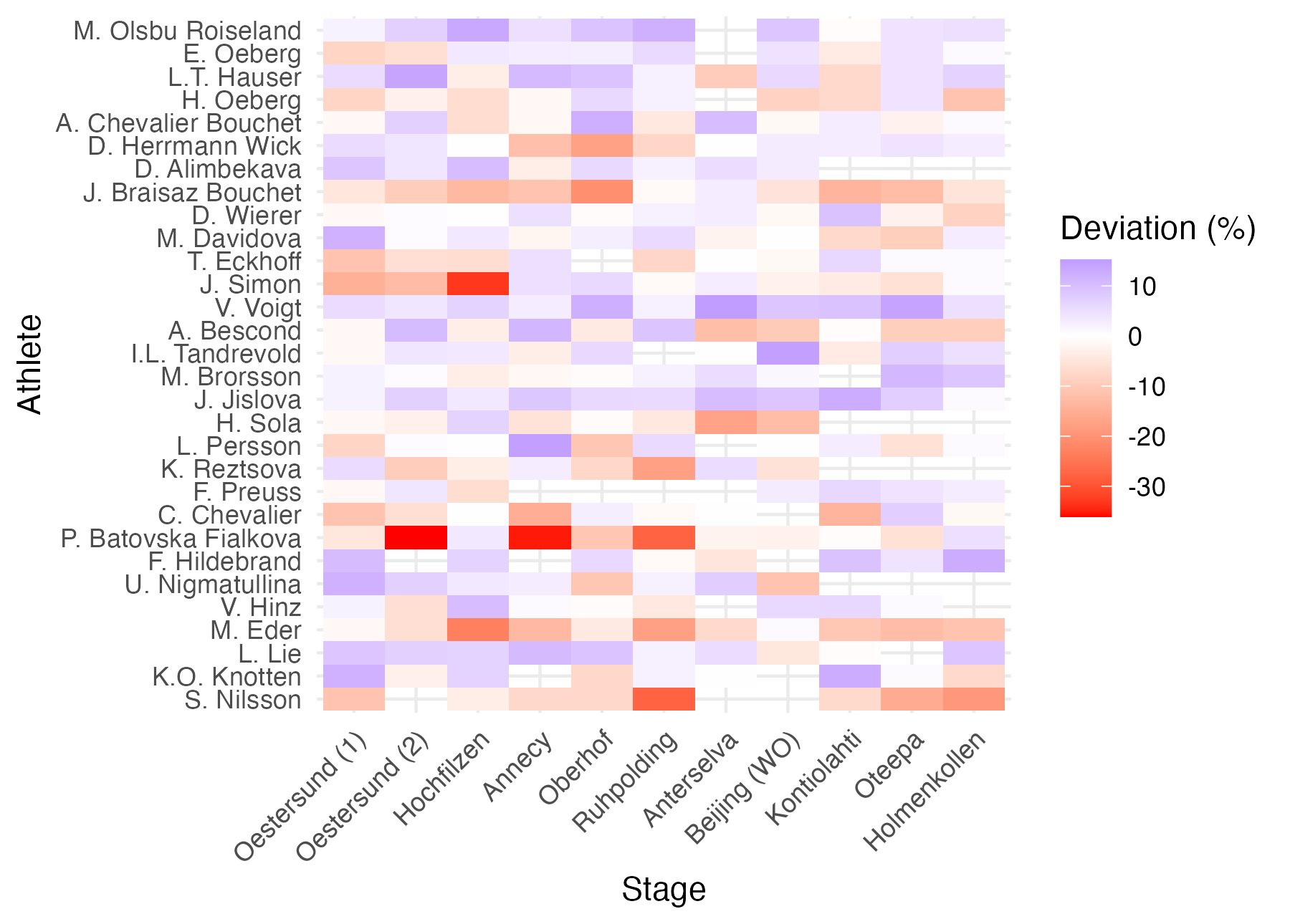}
   \caption{Temporal heatmap illustrating athlete-specific shooting patterns across World Cup stages. Each column represents the deviation from the stage-specific average, highlighting how individual athletes' performance compares to the average at each stage.}

    \label{fig:heatmap}
\end{figure}

The temporal heatmap (Figure~\ref{fig:heatmap}) further revealed that shooting patterns varied across stages, emphasizing that these accuracy trends are specific to each athlete. For instance, some athletes maintained consistently high accuracy across all stages, such as Olsbu and Voigt, who performed above the overall average throughout the season. In contrast, athletes like Nilsson displayed consistently lower accuracy across stages. Other athletes demonstrated more varied patterns: Simon started with lower accuracy, improved mid-season, and then slightly declined towards the end, while Herrmann excelled at the beginning and end of the season, with a dip in performance mid-season. Fialkova generally performed below average but had three stages with notably lower scores.

The exploratory analysis highlights the need for models incorporating athlete-specific effects that account for shooting position and race type, as well as dynamic effects for each stage, given that athletes compared differently to the rest of the pack at various points of the season. 

Further investigation using Spearman's rank correlations evaluated the relationship between end-of-season rankings and shooting accuracy across different categories. The overall shooting percentage showed a weak negative correlation with final rankings ($\rho = -0.148$), indicating that higher-ranked athletes generally had better shooting accuracy. This trend was most pronounced in standing shooting ($\rho = -0.231$) and, to a lesser extent, in pursuit races ($\rho = -0.137$). Individual shooting metrics like prone ($\rho = -0.079$), individual races ($\rho = -0.0371$), and sprint ($\rho = -0.137$) also demonstrated similar weak associations with rank, suggesting varied influences of shooting types and race formats on performance.

\subsection{Model Insights}

In this section, we investigate the insights derived from our Bayesian hierarchical model. The model parameters are interpreted to understand the various factors that influence shooting performance in biathlon competitions. We focus on four main parameter classes: stage effects ($\mu$ parameters), dynamic athlete-specific effects ($\beta$ parameters), position-specific athlete effects ($\gamma$ parameters), and race-specific athlete effects ($\omega$ parameters).

\begin{figure}
    \centering
    \includegraphics[width=0.75\textwidth]{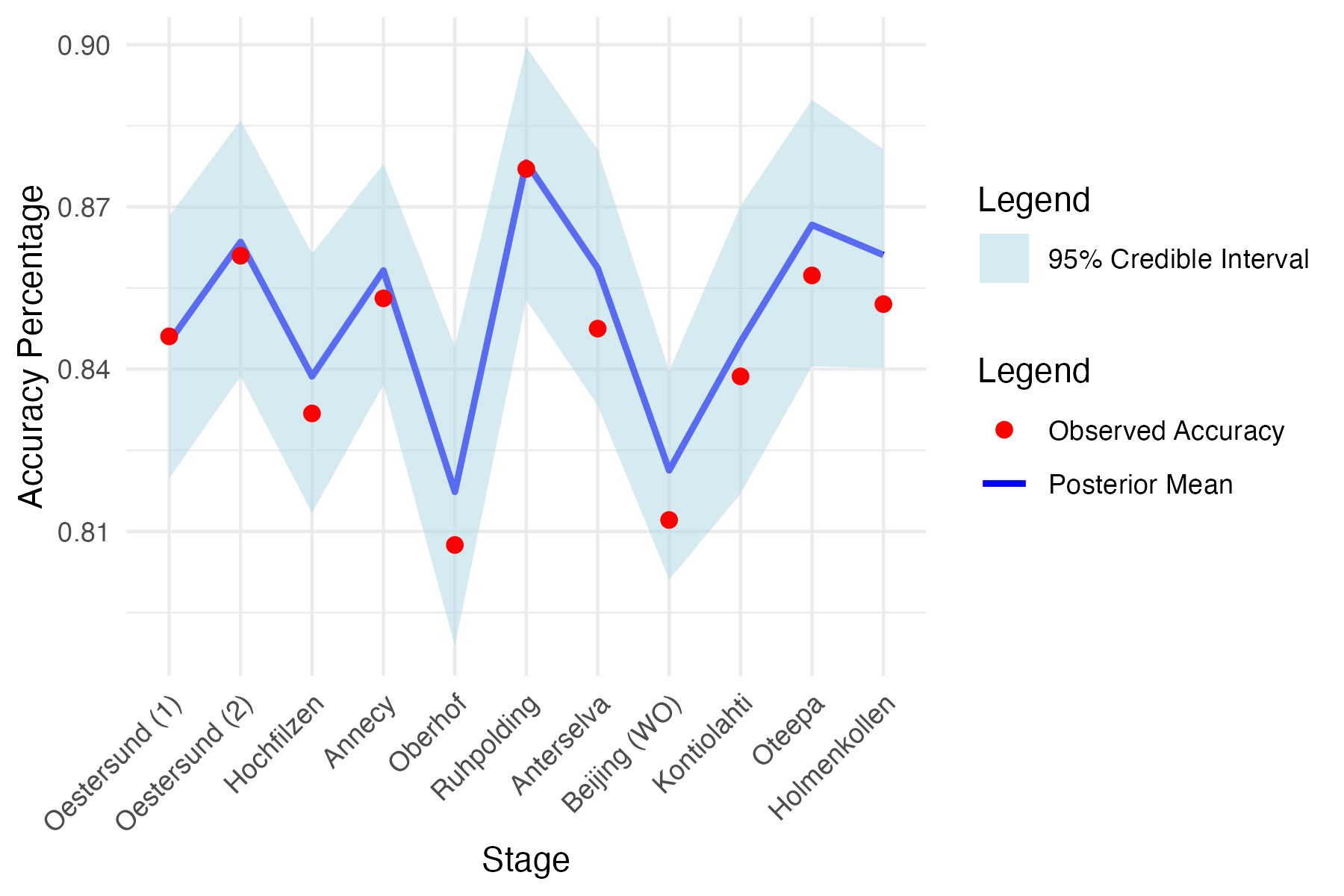}
    \caption{Posterior means and 95\% credible intervals of the $\mu$ parameters across the eleven World Cup stages, representing baseline shooting performance. The observed accuracy percentages are overlaid as red points.}
    \label{fig:mu}
\end{figure}

\begin{figure}
    \centering
    \includegraphics[width=0.85\textwidth]{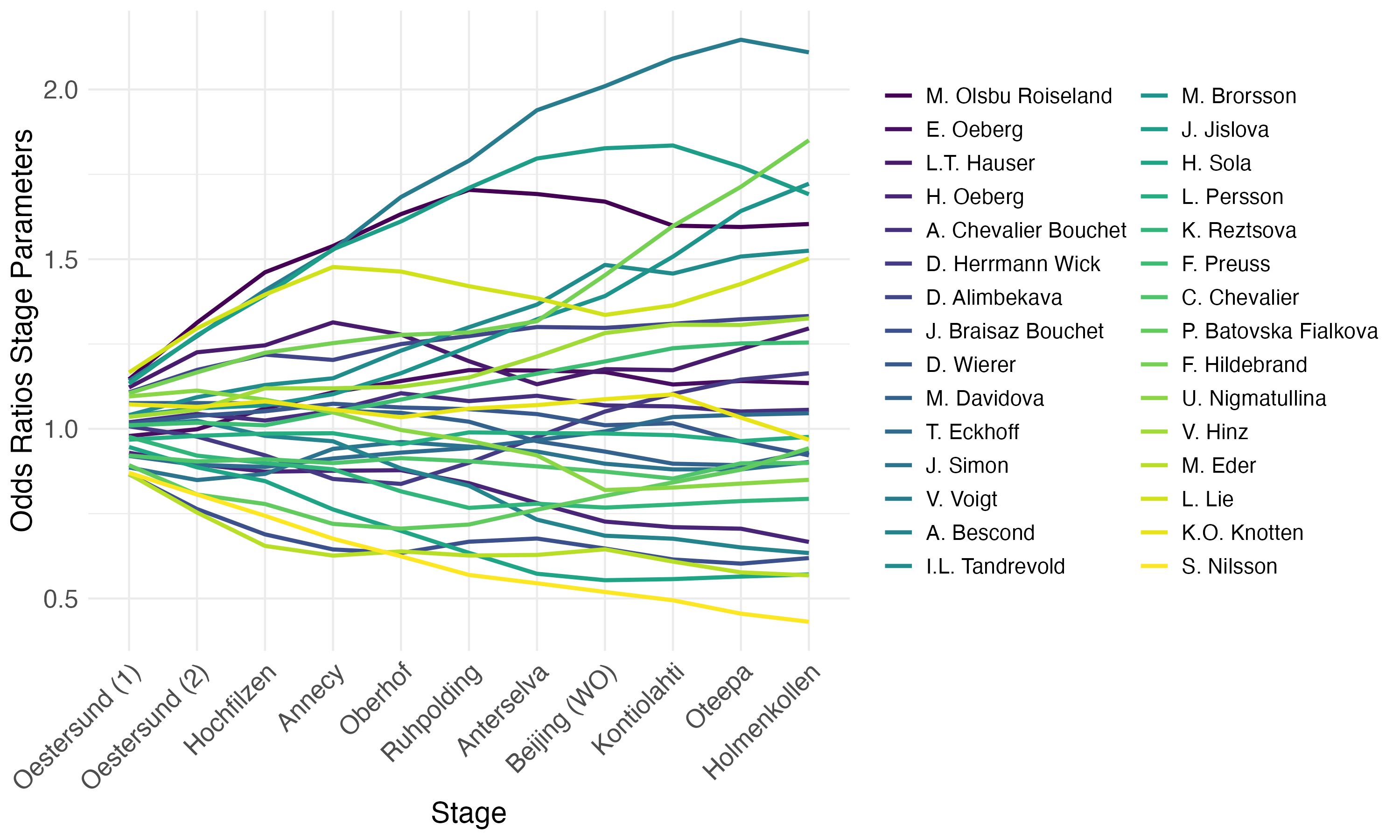}
    \caption{Athlete-specific posterior means of the $\beta$ parameters across the eleven stages, showing how each athlete's performance compares to the average of the group. }
    \label{fig:beta}
\end{figure}

The $\mu$ parameters reported in Figure \ref{fig:mu} effectively capture baseline shooting performance across the World Cup stages, with the posterior means aligning well with observed accuracy percentages for most stages. This demonstrates the model's ability to reflect key performance trends accurately. However, there are two notable instances where the model slightly overestimates accuracy: at Oberhof, a notoriously challenging track due to its steep hill and unpredictable wind conditions, and at the Olympic Games, where athletes face heightened pressure and different snow conditions. Despite these minor discrepancies, the model performs well overall, capturing the nuances of stage-specific performance variations.

\begin{figure}
    \centering
    \includegraphics[width=0.4\textwidth]{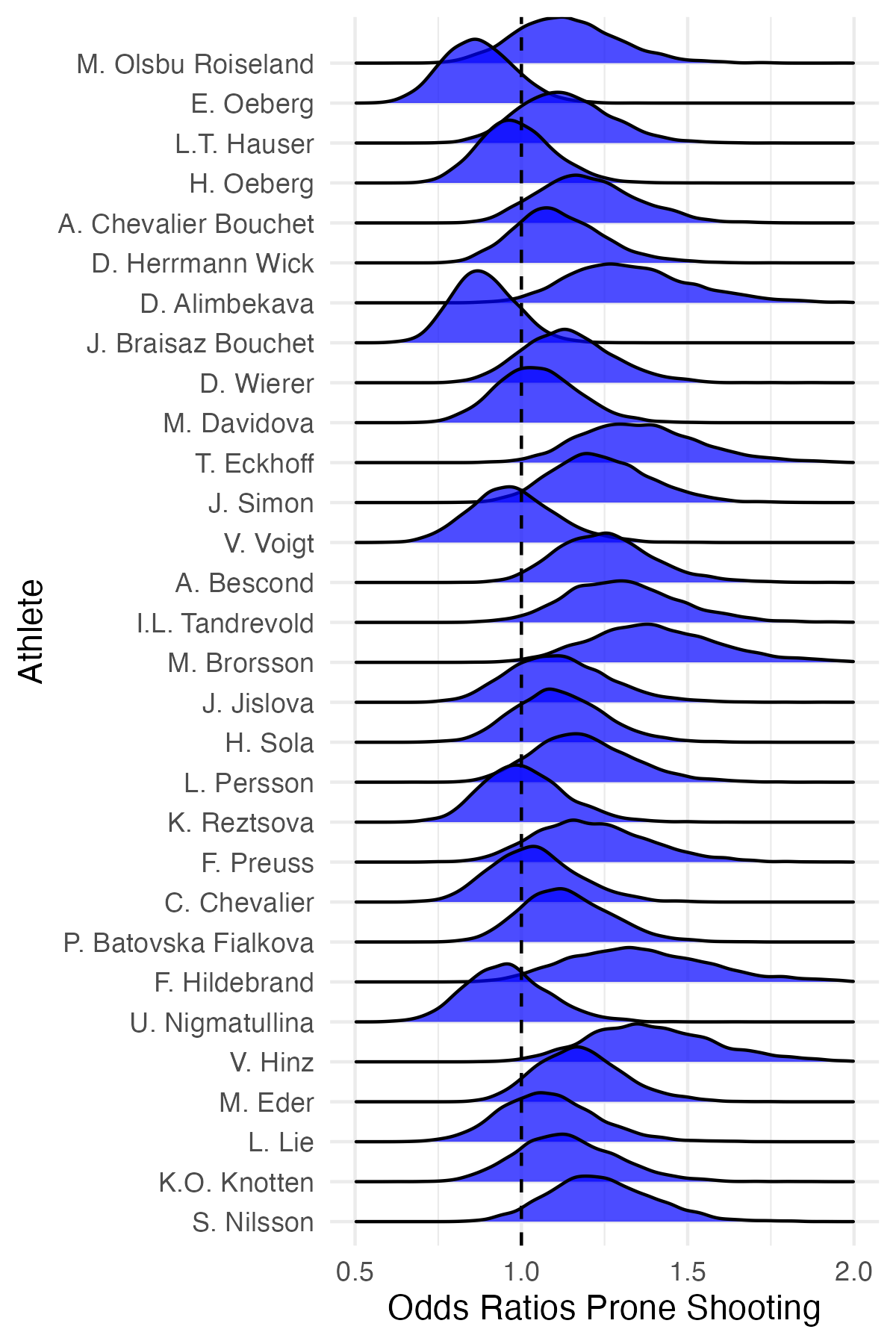}
    \caption{Ridge plot showing the posterior distributions of the prone shooting parameters for each athlete.}
    \label{fig:ridge}
\end{figure}

The $\beta$ parameters illustrated in Figure \ref{fig:beta} show how each athlete's shooting performance at each stage compares to the other 29 athletes, while accounting for prior shooting accuracies through a Markov process. The figure reveals that athletes have unique and often non-monotonic performance trajectories over the season. For example, two athletes depicted in bright green stand out. Hildebrand, with odds ratios generally above one, indicates that she is shooting better than the average of the group, and her performance notably improves towards the end of the season. Conversely, Fialkova Batovska, with odds ratios below one, performs below the average for most of the season but shows a steady improvement, narrowing the gap as the season progresses.

One notable observation from Figure \ref{fig:beta} is that there is no clear pattern indicating that better-ranked athletes (represented by darker colors) have higher $\beta$ parameters. This suggests that overall rankings are not consistently aligned with stronger shooting performance across all stages, highlighting the complexity and variability in athlete performance throughout the season.

The ridge plot in Figure \ref{fig:ridge} displays the posterior distributions of the prone shooting parameters for each athlete, illustrating how their prone shooting strength compares to their standing performance. Since these parameters sum to one for each athlete, there is no need to show the standing shooting parameters separately. The plot reveals distinct patterns: for instance, Oeberg and Braisaz-Bouchet exhibit a clear weakness in prone shooting, with distributions centered well below one, indicating stronger performance in the standing position. In contrast, athletes like Reztsova have distributions centered around one, reflecting similar shooting performance across both positions. Additionally, some athletes, such as Brorsson, show much wider distributions, indicating less consistent performance and a lack of clear patterns in prone shooting accuracy.

\begin{figure}
    \centering
    \includegraphics[width=0.91\textwidth]{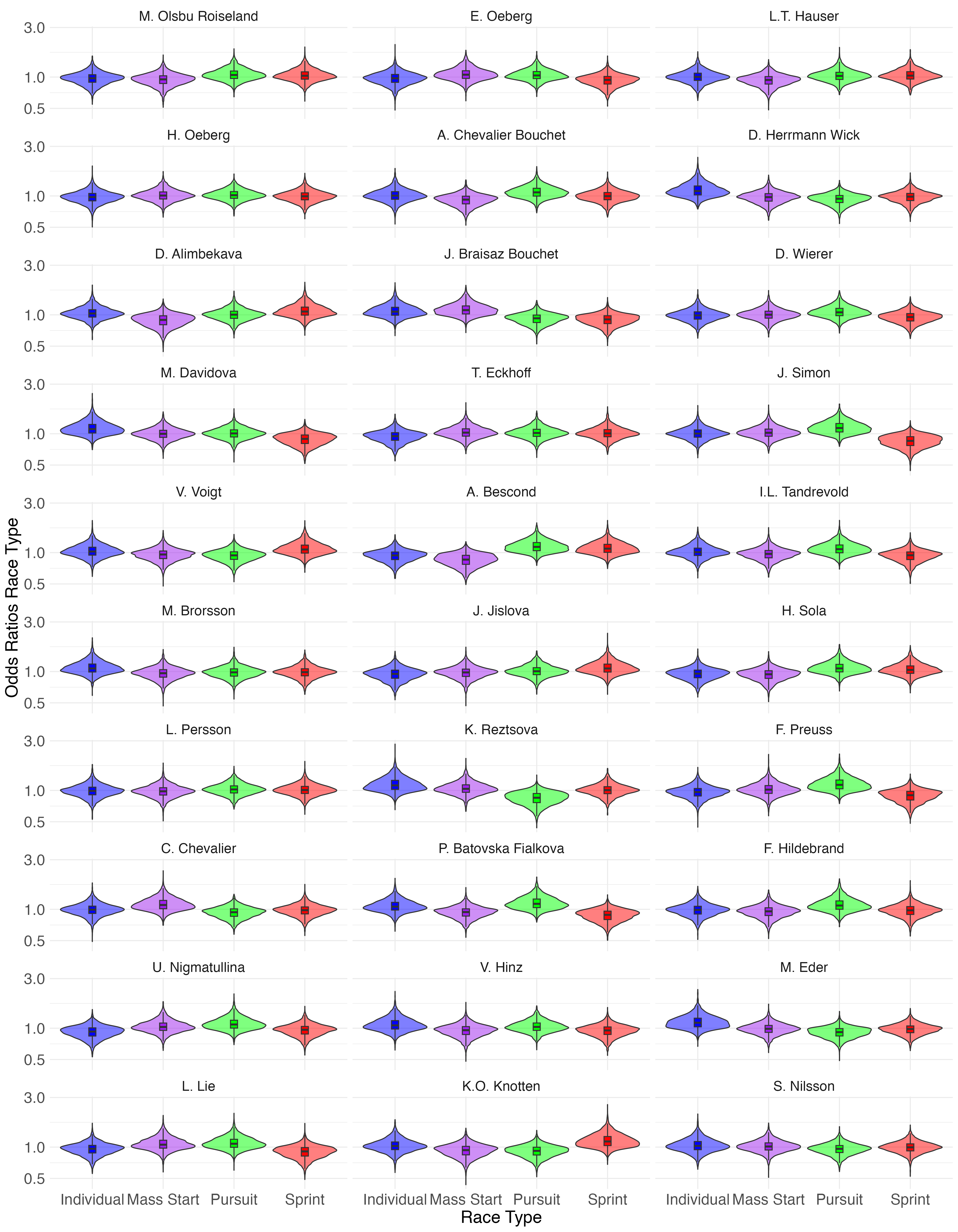}
    \caption{Violin plot showing the distribution of shooting performance parameters for each athlete across different race types. }
    \label{fig:violin}
\end{figure}

The violin plot in Figure \ref{fig:violin} illustrates that each athlete has a distinct shooting profile across different race types. Some athletes, like Olsbu-Roiseland, demonstrate consistent shooting performance across all competitions. In contrast, others specialize in specific race formats. For example, Knotten performs significantly better in Sprint races compared to other formats, Herrmann Wick excels in individual races, and Reztsova shows a noticeable weakness in Pursuit races. This variation underscores the diverse strengths and weaknesses of athletes, shaped by the unique demands of each race type.

\subsection{Predictive Checks}

\begin{figure}
    \centering
    \includegraphics[width=0.8\textwidth]{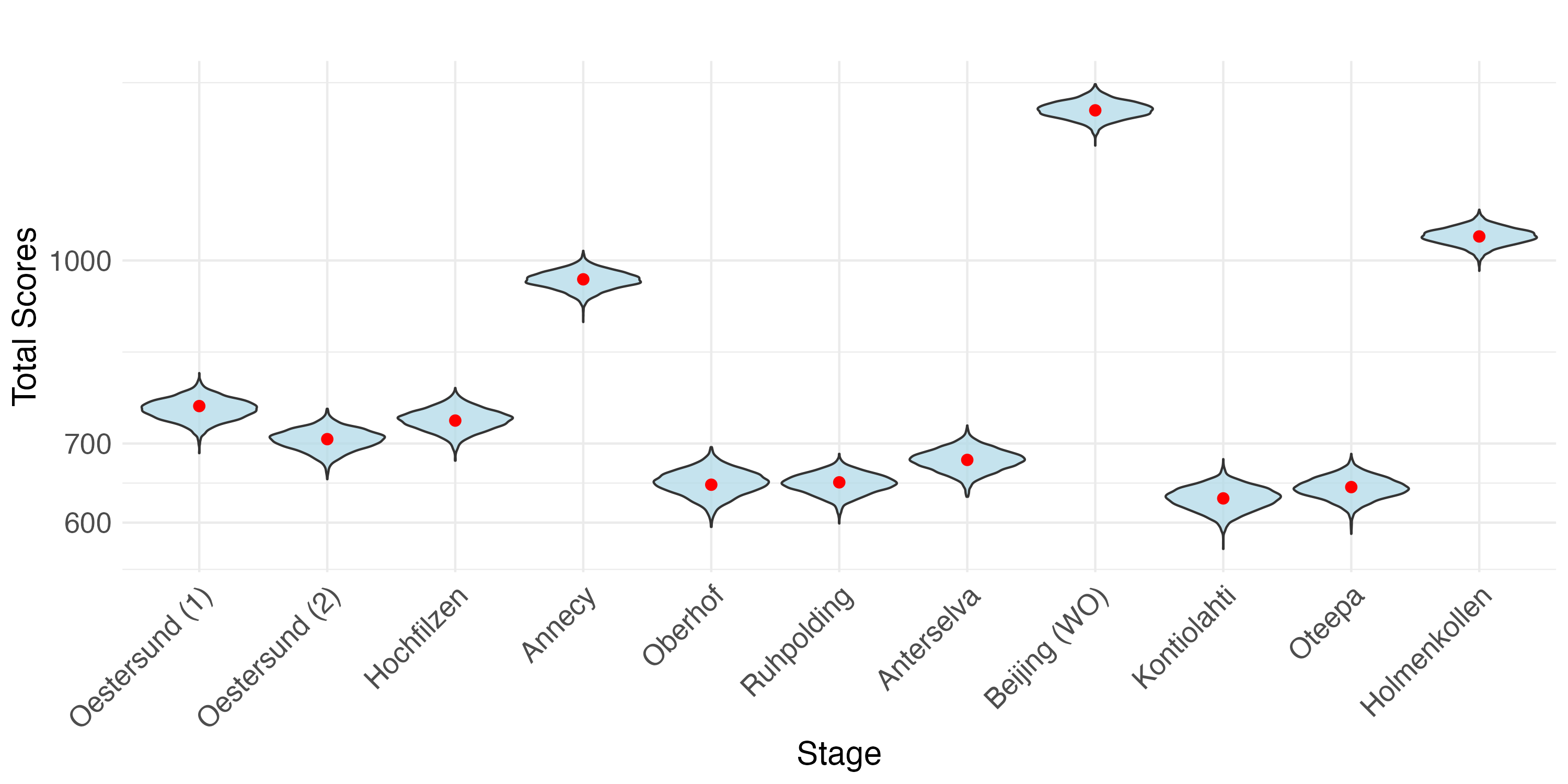}
    \caption{Violin plots displaying the model's predictions for the total number of hits at each stage of the World Cup. Observed totals marked as red dots.}
    \label{fig:predicted}
\end{figure}

The violin plots in Figure \ref{fig:predicted} illustrate the precision of our model's predictions for the total number of hits at each stage of the World Cup. The narrow spread of the violins indicates very accurate predictions overall, with observed values (red dots) aligning closely with the densest parts of the distributions. It is important to note that the total number of hits depends on the number of races held in each stage, which the model captures well.

\begin{figure}
    \centering
    \includegraphics[width=0.8\textwidth]{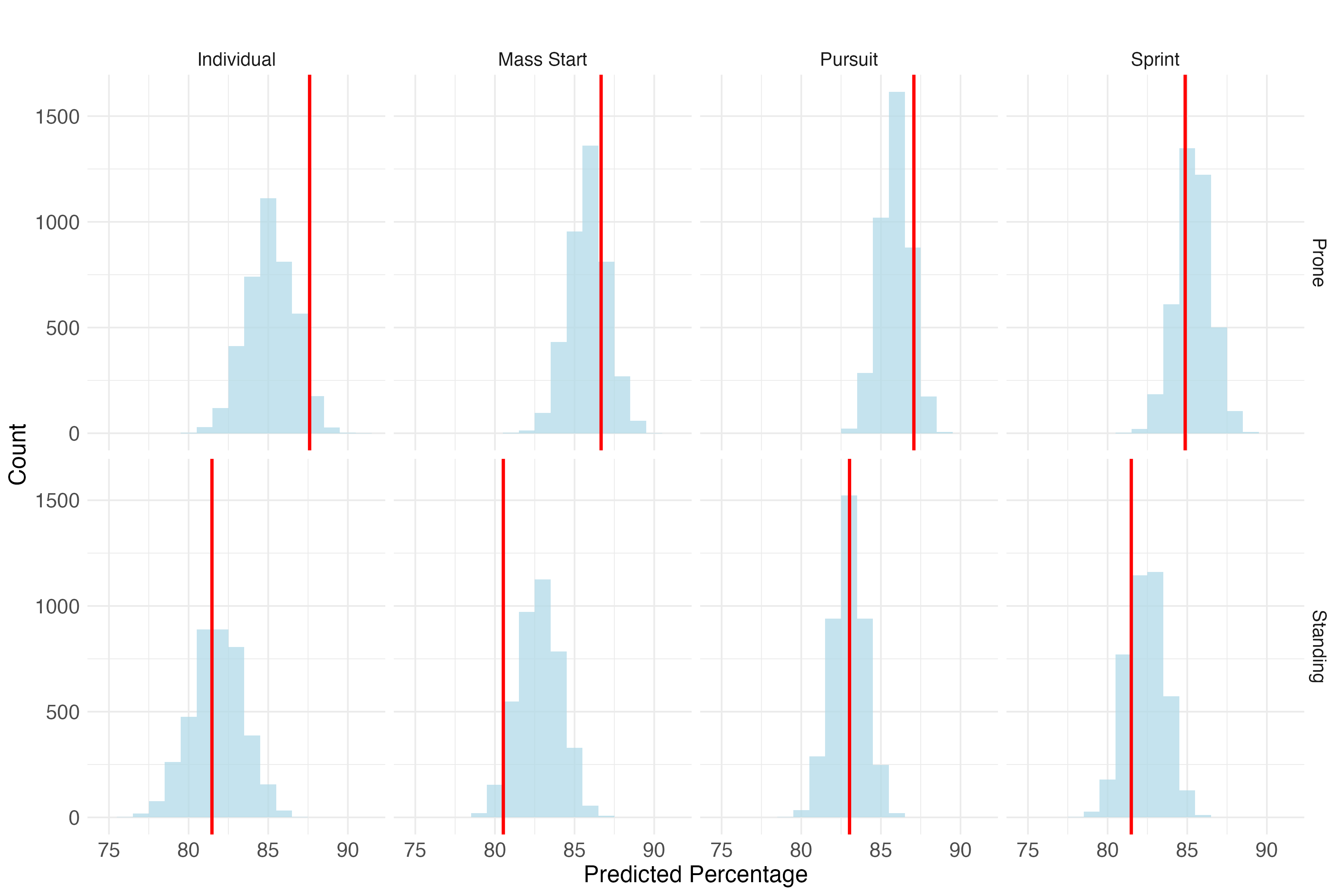}
    \caption{Histograms showing the distribution of predicted shooting percentages across different race types and positions, with vertical red lines indicating the observed shooting percentages.}
    \label{fig:histogram}
\end{figure}

When predictions are broken down by race and shooting position, as shown by the histograms in Figure \ref{fig:histogram}, the model's performance varies. For example, the model tends to underestimate in the prone position for individual races and overestimate in the standing position for mass start races. In other cases, predictions are remarkably accurate. Despite these variations, all observed values lie within the 95\% predictive intervals, demonstrating the model's overall reliability.

\begin{figure}
    \centering
    \includegraphics[width=1\textwidth]{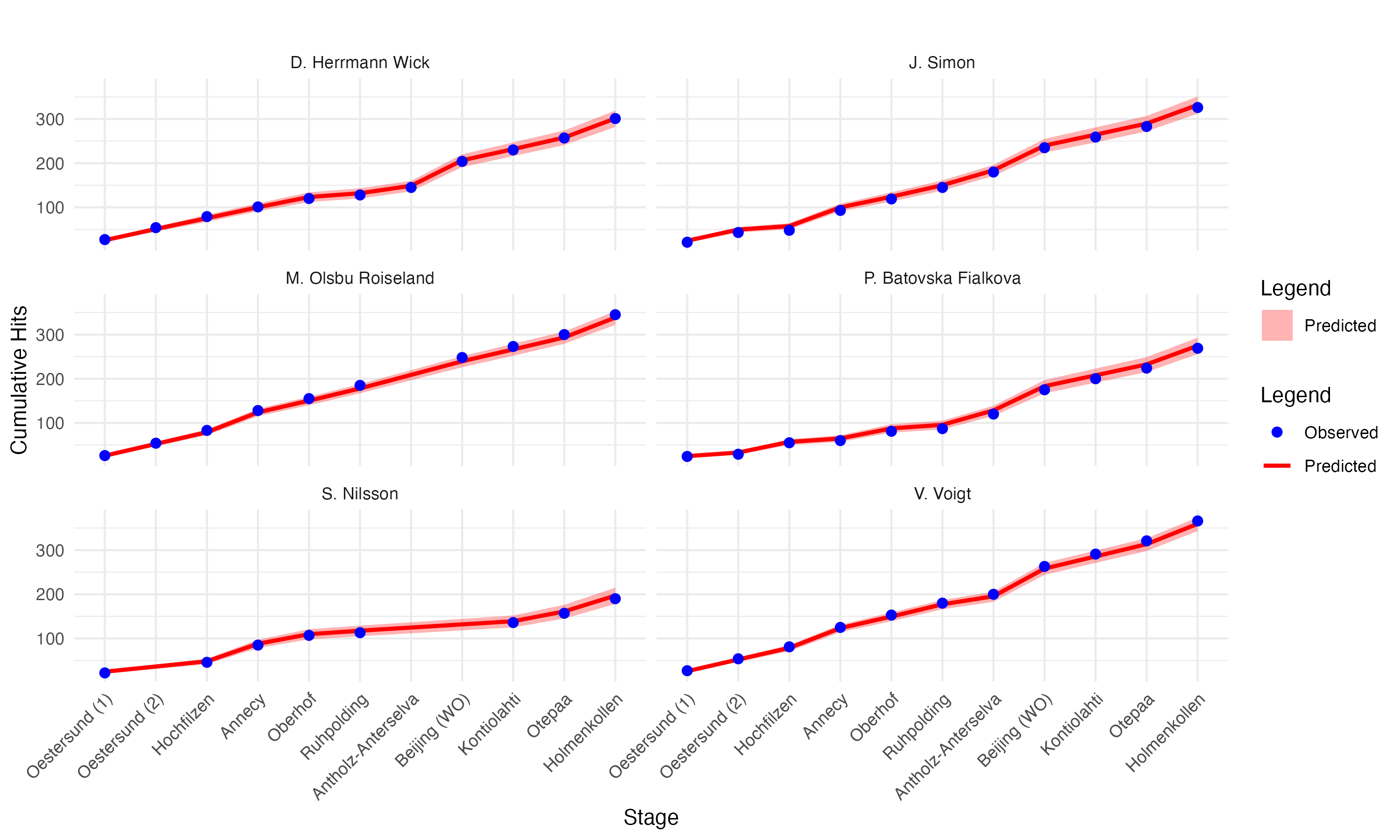}
    \caption{Comparison of the cumulative number of hits throughout the World Cup season for six selected athletes. The red ribbon represents the 95\% predictive interval, while the blue dots indicate the observed cumulative hits. }
    \label{fig:cumulative}
\end{figure}

Finally, Figure \ref{fig:cumulative} showcases individual-specific predictions for six representative athletes, presenting the cumulative number of hits throughout the season. The model delivers highly accurate predictions, even for athletes with non-regular performances, such as Batovska Fialkova. It also handles cases with limited data, like Nilsson, who missed several races, maintaining impressive predictive accuracy under these challenging conditions.

\section{Discussion}

\subsection{Summary and Discussion of Findings}

Our analysis confirms the presence of contextual factors affecting shooting performance, notably highlighted by the low shooting accuracy observed at the Olympic Games (Figure \ref{fig:mu}). The unique pressure of competing for medals awarded only once every four years seems to impact athlete performance. This finding aligns with prior studies that emphasize how heightened stress and environmental changes, such as differing snow conditions, influence shooting accuracy \citep{heinrich2021selection, harb2019choking}.

The exploratory analysis revealed a relatively weak association between shooting percentages and final rankings. Although higher-ranked athletes generally had better shooting accuracy (Table \ref{table1}), the correlation was modest, as discussed in previous literature \citep{maier2018predicting}. However, our model’s $\beta$ parameters (Figure \ref{fig:beta}) do not strongly reinforce this trend. Instead, they display mixed and non-monotonic performance patterns, indicating that athlete-specific strengths vary throughout the season. This variability suggests that shooting performance is only one of many contributing factors to an athlete’s overall success.

Our findings also reaffirm the position-specific effects on shooting accuracy. Prone shooting generally exhibits higher hit rates compared to standing (Table \ref{table}), consistent with previous research attributing this difference to greater rifle stability and reduced body sway in prone shooting \citep{sattlecker2014,sattlecker2017}. Nonetheless, our analysis highlights that these effects are athlete-specific. For instance, some athletes demonstrate a notable strength in standing shooting, while others struggle significantly in this position, as shown in Figure \ref{fig:ridge}. This observation confirms that shooting performance is highly individualized.

Regarding race format effects, we found that the influence of race type on shooting accuracy is smaller compared to the position effect. Interestingly, the highest shooting percentages in our dataset were recorded in pursuit races, contrary to the trends reported in previous studies, which identified individual races as having higher hit rates \citep{skattebo2018variability}. This discrepancy underscores the variability in shooting performance across different race types, further supported by the unique race-specific profiles observed in Figure \ref{fig:violin}. Athletes such as Knotten and Herrmann Wick clearly excel in certain race formats while struggling in others, emphasizing that race type effects are also athlete-dependent.

\subsection{Model Predictive Ability}

Our Bayesian modeling approach provided a flexible framework for simulating outcomes and assessing the quality of the model across multiple levels. This approach not only enabled us to evaluate overall predictive performance but also to scrutinize predictions at an individual athlete level. The predictive checks demonstrated the model's ability to capture key aspects of shooting performance, including skills-dependent, stage-dependent, and athlete-specific characteristics. Overall, the Bayesian framework provided a comprehensive and probabilistically grounded approach to modeling shooting performance, offering valuable insights into how skills and contextual factors influence outcomes throughout the biathlon season.

\subsection{Limitations}

Our study provides valuable insights into biathlon shooting performance using a Bayesian hierarchical model, but several limitations must be acknowledged, with implications for data collection and model development.

First, our analysis focused on a single season of data from the 2021/22 Women’s World Cup. While this approach offered a comprehensive view of shooting performance for that season, it is unclear whether the athlete-specific effects observed represent stable, long-term skills or are influenced by seasonal variability. Extending this analysis to multiple seasons could help determine the persistence of these effects. However, this would introduce complications, as athletes do not consistently remain in the top 30, may skip races, or take career breaks.

Next, our model does not address potential differences between male and female biathletes. Prior research, such as that by \citet{heinrich2021selection}, has indicated that gender-specific factors, such as physiological and psychological responses to competition, can significantly impact performance. Analyzing whether the patterns observed in our model are consistent for male biathletes or if there are notable gender differences would enrich our understanding of performance disparities between genders.

Additionally, our model assumes independence among shots within each five-shot sequence, following a binomial distribution. However, this assumption may be overly simplistic, as research by \citet{maier2018predicting} suggests interdependencies between shots, where the order of shots and the outcomes of preceding shots may influence subsequent accuracy. Modeling these dependencies would require more complex statistical frameworks and detailed data capturing the sequence and context of each shot. Moreover, our model does not incorporate dynamic physiological or psychological states that could change over time, as highlighted by \citet{glickman2015stochastic}. Addressing these limitations would necessitate continuous biometric monitoring and data on psychological stressors, which are often challenging to obtain and may not be available for all athletes.

Overall, addressing these limitations would demand not only more sophisticated modeling techniques but also access to high-resolution, context-rich data, which is often difficult to collect and share widely.

\subsection{Advantages and Limitations of the Bayesian Approach}

Bayesian hierarchical models provide several key advantages for sports performance analysis. First and foremost, their interpretability makes complex parameter estimates accessible to non-statisticians, such as coaches and performance analysts. Each parameter - whether related to shooting accuracy, race effects, or athlete-specific skills - has a clear, intuitive meaning, offering actionable insights for coaching and performance optimization.

Another strength lies in the Bayesian framework's ability to explicitly quantify uncertainty. This is critical in high-stakes sports like biathlon, where understanding risks can directly influence decision-making and strategy. Predictive distributions, derived from posterior samples, allow for nuanced assessments of likely outcomes and confidence levels, facilitating more informed interventions.

Despite these benefits, Bayesian models come with limitations, primarily related to computational demands. Estimating parameters requires specialized software and significant resources, which can be a drawback compared to simpler models. Nevertheless, with tools like \texttt{R-JAGS}, our model can achieve efficient computation even with large datasets. Furthermore, while machine learning approaches may sometimes yield higher predictive accuracy, they lack the transparency offered by Bayesian models, making them less suitable for contexts where understanding underlying factors is essential \citep{rudin2019stop}.

\subsection{Conclusions}

Our analysis emphasizes the value of incorporating athlete-specific and time-varying effects to capture the complexities of shooting performance in biathlon. The Bayesian hierarchical model used in this study reveals how performance evolves over the season and provides probabilistic insights that are both interpretable and actionable. By leveraging this model, coaches can design tailored training programs that address specific weaknesses, such as an athlete's struggles with standing shooting under pressure.

One of the standout features of the Bayesian approach is its balance between interpretability and flexibility. While more computationally intensive than simpler models, the framework's capacity to model dynamic, stage-to-stage variations offers a detailed understanding of performance patterns. Although deep learning models may promise slight improvements in predictive accuracy, they often fall short in terms of explainability, which is crucial for practical applications in sports coaching.

Overall, this work demonstrates the practical utility of Bayesian hierarchical modeling in sports analytics, particularly for biathlon. By providing a clear, data-driven foundation, this approach not only enhances current training practices but also sets the stage for future research in performance modeling, contributing to a deeper understanding of both shooting and broader athletic dynamics in winter sports.

\bibliography{references}

\end{document}